\documentclass[12pt]{article}
\usepackage{amsmath}
\usepackage{amsfonts}
\usepackage{amssymb}
\usepackage{cite}

\catcode`\@=11
\def\numberbysection{\@addtoreset{equation}{section}
        \def\theequation{\thesection.\arabic{equation}}}
\numberbysection

\newcommand{\be}{\begin{equation}}
\newcommand{\ee}{\end{equation}}
\newcommand{\ben}{\begin{eqnarray}}
\newcommand{\een}{\end{eqnarray}}

\begin{document}

\newlength{\lno} \lno1.5cm \newlength{\len} \len=\textwidth%
\addtolength{\len}{-\lno}

\setcounter{page}{0}

\baselineskip7mm
\renewcommand{\thefootnote}{\fnsymbol{footnote}}
\newpage %
\setcounter{page}{0}

\begin{titlepage}
\vspace{0.5cm}
\begin{center}
{\Large\bf Algebraic Bethe ansatz for the six vertex model with upper triangular $K$-matrices}\\
\vspace{1cm}
{\large  R.A. Pimenta\footnote{pimenta@df.ufscar.br} and A. Lima-Santos\footnote{dals@df.ufscar.br}} \\
\vspace{1cm}
{\large \em Universidade Federal de S\~ao Carlos, Departamento de F\'{\i}sica \\
Caixa Postal 676, CEP 13569-905,~~S\~ao Carlos, Brasil}\\
\end{center}
\vspace{1.2cm}

\begin{abstract}
We consider a formulation of the algebraic Bethe ansatz for the six vertex model
with non-diagonal open boundaries. Specifically, we study the case where
both left and right $K$-matrices have an upper triangular form. We show that the main
difficulty entailed by those form of the $K$-matrices is the construction
of the excited states. However,  it is possible to treat this problem with aid of
an auxiliary transfer matrix and by means of a generalized creation operator.
\end{abstract}

\vspace{2cm}
\begin{center}
\end{center}
\vfill
\begin{center}
\small{\today}
\end{center}
\end{titlepage}

\baselineskip6mm

\section{Introduction}

The introduction of non-periodic boundary conditions into the framework of
the quantum inverse scattering method ({\small QISM}) was performed by
Sklyanin \cite{Skly} who resorted a new object called $K$-matrix, which
satisfies the reflection equations, according to the Cherednik's work \cite%
{Che}. Together with the $R$-matrix, solution of the Yang-Baxter equation,
the $K$-matrix can be used to construct families of commuting transfer
matrices.

Although the reflection equations admit general solutions, the algebraic
Bethe ansatz ({\small ABA}) has been applied directly only for diagonal $K$%
-matrices. In fact, even multi-state vertex models with diagonal boundaries
have been solved by the ABA since \cite{Skly}. See, for instance, the papers 
\cite{Fan,Guan,LiYueHou,KurakLima,LiShi} and references therein.

In the cases with non-diagonal boundaries, one of the impediments to address
the {\small ABA} is the absence of a simple reference state. For this
reason, it has been frequent the use of alternative methods which does not
rely on the existence of a reference state. For instance, in \cite{Nepo1},
exploring functional relations satisfied by the transfer matrix, Nepomechie
was able to derive the eigenvalues of the open {\small XXZ} chain for
special values of the bulk anisotropy though with a non standard form of the
respective Bethe equations. In subsequent works \cite{Nepo2,MurNepo,Nepoetal}%
, conventional Bethe equations were restored and many situations with
constrained and general values of the boundary parameters were studied. See
also recent developments on this approach \cite{CYSW,Nepo2013}. Another
methods include the separation of variables \cite{FGSW,NFK}, the direct use
of the Yang-Baxter algebra to derive the eigenvalues \cite{Gal} and the
representation theory of the so called $q$-Onsager algebra \cite{BK}. One
disadvantage of the functional methods is in general the lack of information
regarding the eigenvectors of the transfer matrix.

Moreover, results obtained from the {\small ABA} method have also been
reported in literature \cite{cao,Doikou,CRM}. In the work \cite{cao} and its
generalization to the spin-$s$ case \cite{Doikou} convenient local gauge
transformations were used in order to find a reference state as well as to
transform the left and right $K$-matrices into diagonal and upper triangular
matrices respectively. In \cite{CRM}, transformations in both auxiliary and
quantum spaces also map the original {\small XXX}-$s$ spin chain with two
full $K$-matrices into one with one diagonal and one triangular $K$-matrix.
A common feature present in these works is the requirement of constraints in
the boundary parameters.

Recently, the rational six vertex model with two upper triangular boundaries
was solved by Belliard \textit{et al.} \cite{BCR}. In this case, the excited
states do not have a fixed number of magnons and thus the usual {\small ABA}
does not apply,
although the usual reference state is still an eigenvector. Similar settings
were first considered in the coordinate Bethe ansatz setup \cite{CR} and also
in the vertex operator approach \cite{BB}.

The purpose of this work is to present a constructive approach to obtain
generalized excited states, independent of coordinate Bethe ansatz outcomes.
The form of the generalized excited states are fixed by requiring the
vanishing of extra unwanted terms in the {\small ABA} analysis.

This paper is organized as follows. In Section \ref{transfermatrix} we
summarize definitions and relations needed for the {\small ABA} method.
Next, Section \ref{BABA}, we apply the {\small ABA} presenting detailed
calculations for the first, second and third generalized excited states, as
well as the nth generalized Bethe vector. We discuss our results in Section %
\ref{conclu}. We left the appendices for useful relations used in the main
text.

\section{Transfer Matrix}

\label{transfermatrix}

The Sklyanin monodromy and transfer matrix for an open vertex model are
defined by 
\begin{eqnarray}
\mathcal{T}_{a}(u) &=&K_{a}^{+}(u)T_{a}(u)K_{a}^{-}(u)T_{a}^{-1}(-u),  \notag
\\
t(u) &=&\text{Tr}_{a}[\mathcal{T}_{a}(u)],  \label{transfer}
\end{eqnarray}%
where $K_{a}^{\pm }(u)$ are the reflection matrices, $T_{a}(u)$ and $%
T_{a}^{-1}(-u)$ are the monodromy matrices associated to a chain of length $%
L $ given by a ordered product of $R$-matrices 
\begin{equation}
T(u)=R_{a1}(u)\ldots R_{aL}(u),~~T^{-1}(-u)=R_{aL}(u)\ldots R_{a1}(u).
\label{pmm}
\end{equation}%
The products in (\ref{transfer}) and (\ref{pmm}) are performed in an
auxiliary space denoted by $a$ and $n=1,2,\ldots ,L$ refers to a quantum
vector space at the site $n$.

The Yang-Baxter equation 
\begin{equation}
R_{12}(u-v)R_{13}(u)R_{23}(v)=R_{23}(v)R_{13}(u)R_{12}(u-v),  \label{YB}
\end{equation}%
and the reflection equations 
\begin{equation}
R_{12}(u-v)K_{1}^{-}(u)R_{12}(u+v)K_{2}^{-}(v)=K_{2}^{-}(v)R_{12}(u+v)K_{1}^{-}(u)R_{12}(u-v),
\label{RER}
\end{equation}%
\begin{equation}
R_{12}(v-u)K_{1}^{+}(u)^{t_{1}}R_{12}(-u-v-2\eta
)K_{2}^{+}(v)^{t_{2}}=K_{2}^{+}(v)^{t_{2}}R_{12}(-u-v-2\eta
)K_{1}^{+}(u)^{t_{1}}R_{12}(v-u)  \label{REL}
\end{equation}%
guarantee that (\ref{transfer}) commutes for arbitrary spectral parameters, 
\textit{i.e.}, $\left[ t(u),t(v)\right] =0$.

It also follows from (\ref{YB}) to (\ref{REL}) at least two global relations
for the monodromy matrices, namely 
\begin{equation}
\check{R}(u-v)T(u)\otimes T(v)=T(v)\otimes T(u)\check{R}(u-v)
\label{global1}
\end{equation}%
and 
\begin{equation}
R_{12}(u-v)U_{1}(u)R_{12}(u+v)U_{2}(v)=U_{2}(v)R_{12}(u+v)U_{1}(u)R_{12}(u-v),
\label{global2}
\end{equation}%
where $\check{R}(u)=PR(u)$ and $U_{a}(u)=T_{a}(u)K_{a}^{-}(u)T_{a}^{-1}(-u)$.

For the six vertex model the $R$-matrix has the form 
\begin{equation}
R(u)=\left( 
\begin{array}{cccc}
1 &  &  &  \\ 
& b(u) & c(u) &  \\ 
& c(u) & b(u) &  \\ 
&  &  & 1%
\end{array}%
\right) ,  \label{RM}
\end{equation}%
while the upper triangular $K$-matrices \cite{VegaRuiz,GhoZamo} can be
written as 
\begin{equation}
K^{-}(u)=\left( 
\begin{array}{cc}
k_{11}^{-}(u) & k_{12}^{-}(u) \\ 
0 & k_{22}^{-}(u)%
\end{array}%
\right) ,\quad K^{+}(u)=\left( 
\begin{array}{cc}
k_{11}^{+}(u) & k_{12}^{+}(u) \\ 
0 & k_{22}^{+}(u)%
\end{array}%
\right)  \label{KM}
\end{equation}%
where%
\begin{equation}
b(u)=\frac{\sinh (u)}{\sinh (u+\eta )},\quad c(u)=\frac{\sinh (\eta )}{\sinh
(u+\eta )},  \label{trigo1}
\end{equation}%
and%
\begin{eqnarray}
k_{11}^{-}(u) &=&\sinh (u+\xi _{-}),\quad k_{12}^{-}(u)=\beta _{-}\sinh
(2u),\quad  \notag \\
k_{22}^{-}(u) &=&\sinh (\xi _{-}-u),  \notag \\
k_{11}^{+}(u) &=&\sinh (-u-\eta +\xi _{+}),\quad k_{12}^{+}(u)=\beta
_{+}\sinh (-2u-2\eta ),\quad  \notag \\
k_{22}^{+}(u) &=&\sinh (u+\eta +\xi _{+}).  \label{trigo2}
\end{eqnarray}
In the rational limit, these functions reduce to%
\begin{equation}
b(u)=\frac{u}{u+\eta },\quad c(u)=\frac{\eta }{u+\eta }  \label{rational1}
\end{equation}%
and%
\begin{eqnarray}
k_{11}^{-}(u) &=&u+\xi _{-},\quad k_{12}^{-}(u)=2\beta _{-}u,  \notag \\
k_{22}^{-}(u) &=&\xi _{-}-u,  \notag \\
k_{11}^{+}(u) &=&-u-\eta +\xi _{+},\quad k_{12}^{+}(u)=-2\beta _{+}(u+\eta
),\quad  \notag  \label{rational2} \\
k_{22}^{+}(u) &=&u+\eta +\xi _{+}.
\end{eqnarray}%
In addition to the spectral parameter $u$ we have $\eta $ which parametrizes
the anisotropy and $\xi _{\pm }$, $\beta _{\pm }$ are the four free
parameters characterizing the boundaries. By taking the first derivative
of the transfer matrix (\ref{transfer1}) we can obtain the corresponding
{\small XXZ} Hamiltonian with non-diagonal boundary terms \cite{Skly,VegaRuiz},
\begin{eqnarray}
 H&=&\sum_{n=1}^{L-1}
\left[\sigma_{n}^x\sigma_{n+1}^x+\sigma_{n}^y\sigma_{n+1}^y+\cosh(\eta)\sigma_{n}^z\sigma_{n+1}^z\right]
\nonumber\\&-&
\frac{\sinh(\eta)}{\sinh(\xi_{+})}\left[\beta_{+}(\sigma_1^x+i\sigma_1^y)+\cosh(\xi_{+})\sigma_1^z\right]+
\frac{\sinh(\eta)}{\sinh(\xi_{-})}\left[\beta_{-}(\sigma_L^x+i\sigma_L^y)+\cosh(\xi_{-})\sigma_L^z\right],
\nonumber\\
\end{eqnarray}
where $\sigma_n^{x,y,z}$ are the standard Pauli matrices acting on the site $n$.

The monodromy matrix $U_{a}(u)=T_{a}(u)K_{a}^{-}(u)T_{a}^{-1}(-u)$ can be
represented by a $2\times 2$ matrix 
\begin{equation}
U_{a}(u)=\left( 
\begin{array}{cc}
\mathcal{A}(u) & \mathcal{B}(u) \\ 
\mathcal{C}(u) & \mathcal{D}(u)%
\end{array}%
\right),  \label{monoU}
\end{equation}%
where $\mathcal{A}(u)$, $\mathcal{B}(u)$, $\mathcal{C}(u)$ and $\mathcal{D}%
(u)$ are operators on the Hilbert space $\otimes_{i=1}^L\mathbb{C}^2$. These
operators satisfy commutation relations thanks to (\ref{global2}). The four
relevant relations for this work are, 
\begin{equation}
\mathcal{B}(u)\mathcal{B}(v)=\mathcal{B}(v)\mathcal{B}(u),  \label{commutBB}
\end{equation}%
\begin{equation}
\mathcal{A}(u)\mathcal{B}(v)=a_{1}(u,v)\mathcal{B}(v)\mathcal{A}%
(u)+a_{2}(u,v)\mathcal{B}(u)\mathcal{A}(v)+a_{3}(u,v)\mathcal{B}(u)\tilde{%
\mathcal{D}}(v),  \label{commutAB}
\end{equation}%
\begin{equation}
\tilde{\mathcal{D}}(u)\mathcal{B}(v)=b_{1}(u,v)\mathcal{B}(v)\tilde{\mathcal{%
D}}(u)+b_{2}(u,v)\mathcal{B}(u)\tilde{\mathcal{D}}(v)+b_{3}(u,v)\mathcal{B}%
(u)\mathcal{A}(v),  \label{commutDB}
\end{equation}%
\begin{eqnarray}
\mathcal{C}(u)\mathcal{B}(v) &=&c_{1}(u,v)\mathcal{B}(v)\mathcal{C}%
(u)+c_{2}(u,v)\mathcal{A}(v)\mathcal{A}(u)+c_{3}(u,v)\mathcal{A}(u)\mathcal{A%
}(v)  \notag \\
&+&c_{4}(u,v)\mathcal{A}(v)\tilde{\mathcal{D}}(u)+c_{5}(u,v)\mathcal{A}(u)%
\tilde{\mathcal{D}}(v)+c_{6}(u,v)\tilde{\mathcal{D}}(u)\mathcal{A}(v)  \notag
\\
&+&c_{7}(u,v)\tilde{\mathcal{D}}(u)\tilde{\mathcal{D}}(v),  \label{commutCB}
\end{eqnarray}%
where we have used (\ref{global1}) to define $\tilde{\mathcal{D}}(u)=%
\mathcal{D}(u)-f(u)\mathcal{A}(u)$ with $f(u)=c(2u)$. The explicit
expressions of the coefficients $a_{j}(u,v)$, $b_{j}(u,v)$ and $c_{j}(u,v)$
are given in Appendix \ref{commutcoef}.

We next use these relations in order to diagonalize the transfer matrix (\ref%
{transfer}).

\section{Bethe ansatz analysis}

\label{BABA}

Taking into account the representation (\ref{monoU}) as well as the upper
triangular form of the left boundary matrix (\ref{KM}), the transfer matrix
has the form 
\begin{eqnarray}
t(u) &=&k_{11}^{+}(u)\mathcal{A}(u)+k_{22}^{+}(u)\mathcal{D}(u)
+k_{12}^{+}(u)\mathcal{C}(u)  \notag \\
&=&\omega _{1}(u)\mathcal{A}(u)+\omega _{2}(u)\tilde{\mathcal{D}}%
(u)+k_{12}^{+}(u)\mathcal{C}(u),  \label{transfer1}
\end{eqnarray}%
where 
\begin{eqnarray}
\omega _{1}(u) &=&k_{11}^{+}(u)+f(u)k_{22}^{+}(u),~~  \notag \\
\omega _{2}(u) &=&k_{22}^{+}(u).  \label{omega}
\end{eqnarray}

The presence of the annihilation operator $\mathcal{C}(u)$ in (\ref%
{transfer1}) hinders the usual task of finding its eigenvectors. In fact,
the excited states of the periodic or diagonal boundary models are usually
constructed by applying the $\mathcal{B}$-operators to the reference state
as a consequence of the commutativity of the transfer matrix and the total
spin operator $\sum_{i=1}^{L}S_{i}^{z}$. For the upper triangular $K$%
-matrices this is no longer true and the excited states have to be
constructed in another way.

\subsection{The reference state}

An important step in the {\small ABA} technique is the choice of a reference
state from which the excited states are constructed. It turns out that the
state 
\begin{equation}
\Psi _{0}=\left( 
\begin{array}{c}
1 \\ 
0%
\end{array}%
\right) _{(1)}\otimes \left( 
\begin{array}{c}
1 \\ 
0%
\end{array}%
\right) _{(2)}\otimes \cdots \otimes \left( 
\begin{array}{c}
1 \\ 
0%
\end{array}%
\right) _{(L)}  \label{refstate}
\end{equation}%
is a eigenstate of (\ref{transfer1}). This is a consequence of the structure
of the $U_{a}(u)$ matrix elements when the $K^{-}(u)$ matrix has the form (%
\ref{KM}). In fact, with the help of (\ref{global1}), it is not difficult to
calculate 
\begin{equation}
\mathcal{A}(u)\Psi _{0}=\Delta _{1}(u)\Psi _{0},~~\tilde{\mathcal{D}}(u)\Psi
_{0}=\Delta _{2}(u)\Psi _{0},~~\mathcal{C}(u)\Psi _{0}=0,~~\mathcal{B}%
(u)\Psi _{0}=\ast,  \label{actionrefs}
\end{equation}%
where $\ast $ denotes a state different from $0$\ and $\Psi _{0}$, 
\begin{eqnarray}
\Delta _{1}(u) &=&k_{11}^{-}(u),  \notag \\
~\Delta _{2}(u) &=&\left[ k_{22}^{-}(u)-f(u)k_{11}^{-}(u)\right] b(u)^{2L} .
\label{deltas}
\end{eqnarray}%
Therefore, we have an eigenvalue problem, 
\begin{equation}
t(u)\Psi _{0}=\Lambda _{0}(u)\Psi _{0}  \label{0part}
\end{equation}%
where 
\begin{equation}
\Lambda _{0}(u)=\omega _{1}(u)\Delta _{1}(u)+\omega _{2}(u)\Delta _{2}(u)
\end{equation}%
is the corresponding eigenvalue.

\subsection{The first excited state}

In order to construct the first excited state, we introduce an auxiliary
transfer matrix given by 
\begin{equation}
\bar{t}(u)=\omega _{1}(u)\mathcal{A}(u)+\omega _{2}(u)\tilde{\mathcal{D}}(u).
\label{atransfer}
\end{equation}%
We observe that $t(u)$ and $\bar{t}(u)$ share the same reference state, 
\textit{i.e.}, $t(u)\Psi _{0}=\bar{t}(u)\Psi _{0}=\Lambda _{0}(u)\Psi _{0}$.

The one-particle state of the auxiliary transfer matrix $\bar{t}(u)$ can be
obtained as usual, 
\begin{equation}
\Psi _{1}(u_{1})=\mathcal{B}(u_{1})\Psi _{0},  \label{PSI1}
\end{equation}%
and the action of $\bar{t}(u)$ on the state (\ref{PSI1}), using the
relations (\ref{ABn}) and (\ref{DBn}) of the Appendix \ref{reorder}, is
given by, 
\begin{equation}
\bar{t}(u)\Psi _{1}(u_{1})=\Lambda _{1}(u,u_{1})\Psi _{1}(u_{1})+\left[
\omega _{1}(u)F_{1}(u,u_{1})+\omega _{2}(u)G_{1}(u,u_{1})\right] \mathcal{B}%
(u)\Psi _{0},  \label{tPSI1}
\end{equation}%
where 
\begin{equation}
\Lambda _{1}(u,u_{1})=\omega _{1}(u)\Delta _{1}(u)a_{1}(u,u_{1})+\omega
_{2}(u)\Delta _{2}(u)b_{1}(u,u_{1}).  \label{lam1}
\end{equation}

The form of the upper element of $\mathcal{T}_{a}(u)$, namely $k_{11}^{+}(u)%
\mathcal{B}(u)+k_{12}^{+}(u)\mathcal{D}(u)$, suggests that the first excited
state should contain two terms: one is the usual, $\mathcal{B}(u_{1})\Psi
_{0}$, while another one comes from a diagonal operator actiong on $\Psi_0$.
We propose accordingly the following first excited state\footnote{We remark that we use
the nomenclature ``excited states'' for the eigenvectors of $t(u)$ to distinguish
them from the ``particle states'' of $\bar{t}(u)$, which have a fixed number of magnons.} for $t(u)$, 
\begin{equation}
\Phi _{1}(u_{1})=\Psi _{1}(u_{1})+g(u_{1})\Psi _{0}  \label{ansatzPHI1}
\end{equation}%
where the function $g(u_{1})$ is to be fixed. 

By acting the transfer matrix on $\Phi_1(u_1)$ we find, 
\begin{equation}
t(u)\Phi_1(u_1)=\bar{t}(u)\Psi_1(u_1)+g(u_1)\bar{t}(u)\Psi_0 +k_{12}^{+}(u)%
\mathcal{C}(u)\Psi_1(u_1).
\end{equation}
Now, we use the expression (\ref{tPSI1}) and also (\ref{CBn}) to obtain, 
\begin{eqnarray}
t(u)\Phi _{1}(u_{1}) &=&\Lambda _{1}(u,u_{1})\Phi _{1}(u_{1})+\left[ \omega
_{1}(u)F_{1}(u,u_{1})+\omega _{2}(u)G_{1}(u,u_{1})\right] \mathcal{B}(u)\Psi
_{0}  \notag  \label{tPHI1a} \\
&+&\left\{ g(u_{1})\left[ \Lambda _{0}(u)-\Lambda _{1}(u,u_{1})\right]
+k_{12}^{+}(u)H_{1}(u,u_{1})\right\} \Psi _{0}.
\end{eqnarray}

Besides the usual unwanted term found for the auxiliary transfer matrix (\ref%
{tPSI1}) there is an additional state in (\ref{tPHI1a}). Setting the
coefficient of $\mathcal{B}(u)\Psi _{0}$ equal to zero we obtain, 
\begin{equation}  \label{bethe1particleAUX}
\frac{\Delta _{1}(u_{1})}{\Delta _{2}(u_{1})}= -\frac{a_3(u,u_1)%
\omega_1(u)+b_2(u,u_1)\omega_2(u)} {a_2(u,u_1)\omega_1(u)+b_3(u,u_1)%
\omega_2(u)}.
\end{equation}%
One can verify that the right-hand side of (\ref{bethe1particleAUX}) depends 
\textbf{only} on $u_1$ \cite{Skly}. Thus, we find the Bethe equation for the
first excited in the form, 
\begin{equation}
\frac{\Delta _{1}(u_{1})}{\Delta _{2}(u_{1})}=-\Theta (u_{1}),
\label{bethe1particle}
\end{equation}%
where 
\begin{equation}  \label{thetatrig}
\Theta (u_{1})=\frac{\sinh (2u_{1}+\eta )\sinh (u_{1}+\eta +\xi _{+})}{\sinh
(2u_{1})\sinh (u_{1}-\xi _{+})}.
\end{equation}

On the other hand, the unwanted term proportional to the reference state is
used to extract the expression for $g(u_{1})$, 
\begin{equation}
g(u_{1})=\frac{k_{12}^{+}(u)H_{1}(u,u_{1})}{\Lambda _{1}(u,u_{1})-\Lambda
_{0}(u)}.  \label{expga}
\end{equation}
At a first glance, the right-hand side of (\ref{expga}) is also dependent of
the spectral variable $u$. However, if we take into account the Bethe
equation (\ref{bethe1particle}), identities between the coefficients (\ref%
{commutABcoef}-\ref{commutCBcoef}) and the following relation for the $K$%
-matrix elements, 
\begin{eqnarray}
&&\frac{k_{12}^{+}(u)\omega _{1}(u_{1})}{k_{12}^{+}(u_{1})\left[
a_{2}(u,u_{1})\omega _{1}(u)+b_{3}(u,u_{1})w_{2}(u)\right] }  \notag \\
&=&\frac{1-a_{1}(u,u_{1})}{a_{3}(u,u_{1})\left[ c_{2}(u,u_{1})+c_{3}(u,u_{1})%
\right] -a_{2}(u,u_{1})c_{5}(u,u_{1})}  \label{identity}
\end{eqnarray}%
the expression (\ref{expga}) acquires a simple form, 
\begin{equation}
g(u_{1})=\Delta _{2}(u_{1})\left[ \frac{k_{12}^{+}(u_{1})}{%
k_{11}^{+}(u_{1})+f(u_{1})k_{22}^{+}(u_{1})}\right]  \label{expg}
\end{equation}%
which depends \textbf{only }on $u_{1}$, as expected. The equations (\ref%
{bethe1particle}) to (\ref{expg}) ensure that $\Phi _{1}(u_{1})$ is an
eigenstate of $t(u)$ with eigenvalue (\ref{lam1}).

\subsection{The second excited state}

We proceed in a similar way for the second excited state. First, we consider
the two-particle state of the auxiliary transfer matrix 
\begin{equation}
\Psi _{2}(u_{1},u_{2})=\mathcal{B}(u_{1})\mathcal{B}(u_{2})\Psi _{0},
\label{PSI2}
\end{equation}%
in order to get, 
\begin{eqnarray}
\bar{t}(u)\Psi _{2}(u_{1},u_{2}) &=&\Lambda _{2}(u,u_{1},u_{2})\Psi
_{2}(u_{1},u_{2})  \notag  \label{tPSI2} \\
&+&\left[ \omega _{1}(u)F_{2}(u,u_{1},u_{2})+\omega
_{2}(u)G_{2}(u,u_{1},u_{2})\right] \mathcal{B}(u)\mathcal{B}(u_{1})\Psi _{0}
\notag \\
&+&\left[ \omega _{1}(u)F_{1}(u,u_{1},u_{2})+\omega
_{2}(u)G_{1}(u,u_{1},u_{2})\right] \mathcal{B}(u)\mathcal{B}(u_{2})\Psi _{0}
\end{eqnarray}%
where 
\begin{equation}
\Lambda _{2}(u,u_{1},u_{2})=\omega
_{1}(u)\Delta_1(u)\prod_{j=1}^{2}a_{1}(u,u_{j})+\omega
_{2}(u)\Delta_2(u)\prod_{j=1}^{2}b_{1}(u,u_{j}).  \label{lam2}
\end{equation}

The ansatz for the full second excited state is guessed from the action of $%
k_{11}^{+}(u)\mathcal{B}(u)$ $+k_{12}^{+}(u)\mathcal{D}(u)$ on $\Psi _{0}$
twice, 
\begin{eqnarray}
\Phi _{2}(u_{1},u_{2}) &=&\Psi _{2}(u_{1},u_{2})  \notag \\
&&+g^{(1)}_2(u_{1},u_{2})\Psi _{1}(u_{1})+g^{(1)}_{1}(u_{1},u_{2})\Psi
_{1}(u_{2})  \notag \\
&&+g^{(0)}_{12}(u_{1},u_{2})\Psi _{0}  \label{ansatzPHI2}
\end{eqnarray}%
with the coefficients $g^{(1)}_{1,2}(u_{1},u_{2})$ and $%
g^{(0)}_{12}(u_{1},u_{2})$ to be determined \textit{a posteriori}.

The action of $t(u)$ on the state (\ref{ansatzPHI2}), gathering the previous
results (\ref{tPSI2}), generates many unwanted terms, 
\begin{eqnarray}
t(u)\Phi _{2}(u_{1},u_{2}) &=&\Lambda _{2}(u,u_{1},u_{2})\Phi
_{2}(u_{1},u_{2})  \notag  \label{tPHI2} \\
&+&\left[ \omega _{1}(u)F_{2}(u,u_{1},u_{2})+\omega
_{2}(u)G_{2}(u,u_{1},u_{2})\right] \mathcal{B}(u)\Psi _{1}(u_{1})  \notag \\
&+&\left[ \omega _{1}(u)F_{1}(u,u_{1},u_{2})+\omega
_{2}(u)G_{1}(u,u_{1},u_{2})\right] \mathcal{B}(u)\Psi _{1}(u_{2})  \notag \\
&+&\left\{ g_{2}^{(1)}(u_{1},u_{2})\left[ \omega
_{1}(u)F_{1}(u,u_{1})+\omega _{2}(u)G_{1}(u,u_{1})\right] \right.  \notag \\
&+&\left. g_{1}^{(1)}(u_{1},u_{2})\left[ \omega _{1}(u)F_{1}(u,u_{2})+\omega
_{2}(u)G_{1}(u,u_{2})\right] \right.  \notag \\
&+&\left. k_{12}^{+}(u)H_{21}(u,u_{1},u_{2})\right\} \mathcal{B}(u)\Psi _{0}
\notag \\
&+&\left\{ g_{2}^{(1)}(u_{1},u_{2})\left[ \Lambda _{1}(u,u_{1})-\Lambda
_{2}(u,u_{1},u_{2})\right] +k_{12}^{+}(u)H_{2}(u,u_{1},u_{2})\right\} \Psi
_{1}(u_{1})  \notag \\
&+&\left\{ g_{1}^{(1)}(u_{1},u_{2})\left[ \Lambda _{1}(u,u_{2})-\Lambda
_{2}(u,u_{1},u_{2})\right] +k_{12}^{+}(u)H_{1}(u,u_{1},u_{2})\right\} \Psi
_{1}(u_{2})  \notag \\
&+&\left\{ g_{12}^{(0)}(u_{1},u_{2})\left[ \Lambda _{0}(u)-\Lambda
_{2}(u,u_{1},u_{2})\right] \right.  \notag \\
&+&\left. k_{12}^{+}(u)\left[
g_{2}^{(1)}(u_{1},u_{2})H_{1}(u,u_{1})+g_{1}^{(1)}(u_{1},u_{2})H_{1}(u,u_{2})%
\right] \right\} \Psi _{0}.
\end{eqnarray}

The coefficients of $\mathcal{B}(u)\Psi _{1}(u_{1})$ and $\mathcal{B}(u)\Psi
_{1}(u_{2})$ lead to the Bethe equations, 
\begin{equation}
\frac{\Delta _{1}(u_{1})}{\Delta _{2}(u_{1})}=-\Theta (u_{1})\frac{%
b_{1}(u_{1},u_{2})}{a_{1}(u_{1},u_{2})},\qquad \frac{\Delta _{1}(u_{2})}{%
\Delta _{2}(u_{2})}=-\Theta (u_{2})\frac{b_{1}(u_{2},u_{1})}{%
a_{1}(u_{2},u_{1})},  \label{bethe2particle}
\end{equation}%
while the coefficients of $\Psi_1(u_1)$, $\Psi_1(u_2)$ and $\Psi_0$ give us
the expressions for $g^{(1)}_{1,2}(u_{1},u_{2})$ and $%
g^{(0)}_{12}(u_{1},u_{2})$. Taking into account the Bethe equations (\ref%
{bethe2particle}) we get 
\begin{eqnarray}
g_{1}^{(1)}(u_{1},u_{2}) &=&g(u_{1})p(u_{2},u_{1}),~~  \notag \\
g_{2}^{(1)}(u_{1},u_{2}) &=&g(u_{2})p(u_{1},u_{2}),~~  \notag \\
g_{12}^{(0)}(u_{1},u_{2}) &=&g(u_{1})g(u_{2})q(u_{1},u_{2})
\label{expg1g2g3}
\end{eqnarray}%
where $g(u_{i})$ is given by (\ref{expg}) and we have introduced two new
functions, namely 
\begin{equation}
p(u,v)=b_{1}(u,v)\frac{a_{1}(u,v)}{a_{1}(v,u)},\qquad q(u,v)=\frac{b_{1}(v,u)%
}{a_{1}(u,v)}.  \label{defhk}
\end{equation}

We can check by direct computation that the other unwanted term in (\ref%
{tPSI2}) are automatically null if we take into account the Bethe equations (%
\ref{bethe2particle}) and the expressions (\ref{expg1g2g3}). Thus, the state 
\begin{eqnarray}
\Phi _{2}(u_{1},u_{2}) &=&\Psi _{2}(u_{1},u_{2})  \notag \\
&+&g(u_{2})p(u_{1},u_{2})\Psi _{1}(u_{1})+g(u_{1})p(u_{2},u_{1})\Psi
_{1}(u_{2})  \label{PHI2} \\
&+&g(u_{1})g(u_{2})q(u_{1},u_{2})\Psi _{0}  \notag
\end{eqnarray}%
is an eigenstate of the transfer matrix with energy (\ref{lam2}).

\subsection{The third excited state}

It is expected from the integrability of the model that the second excited
state structure should allow the generalization to find the nth excited
structure. Nevertheless, it was not sufficient to guess the nth excited
state from (\ref{PHI2}). Thus, we also proceed to the the third excited
state.

Following the previous discussions we propose the following structure for
the third excited state, 
\begin{eqnarray}
\Phi _{3}(u_{1},u_{2},u_{3}) &=&\Psi _{3}(u_{1},u_{2},u_{3})
+g_{3}^{(2)}(u_{1},u_{2},u_{3})\Psi _{2}(u_{1},u_{2})  \notag \\
&+& g_{2}^{(2)}(u_{1},u_{2},u_{3})\Psi
_{2}(u_{1},u_{3})+g_{1}^{(2)}(u_{1},u_{2},u_{3})\Psi _{2}(u_{2},u_{3}) 
\notag \\
&+&g_{23}^{(1)}(u_{1},u_{2},u_{3})\Psi
_{1}(u_{1})+g_{13}^{(1)}(u_{1},u_{2},u_{3})\Psi
_{1}(u_{2})+g_{12}^{(1)}(u_{1},u_{2},u_{3})\Psi _{1}(u_{3})  \notag \\
&+&g_{123}^{(0)}(u_{1},u_{2},u_{3})\Psi _{0}  \label{psi3}
\end{eqnarray}%
where the coefficients $g^{(k)}(u_{1},u_{2},u_{3})$ will be determined in
what follows.

As before we first apply the auxiliary transfer matrix $\bar{t}(u)$ to the
three-particle state 
\begin{equation}
\Psi _{3}(u_{1},u_{2},u_{3})=\mathcal{B}(u_{1})\mathcal{B}(u_{2})\mathcal{B}%
(u_{3})\Psi _{0},
\end{equation}%
and, as a result, we obtain, 
\begin{eqnarray}
\bar{t}(u)\Psi _{3}(u_{1},u_{2},u_{3}) &=&\Lambda
_{3}(u,u_{1},u_{2},u_{3})\Psi _{3}(u_{1},u_{2},u_{3})  \notag  \label{tPSI3}
\\
&+&\left[ \omega _{1}(u)F_{3}(u,u_{1},u_{2},u_{3})+\omega
_{2}(u)G_{3}(u,u_{1},u_{2},u_{3})\right] \mathcal{B}(u)\Psi _{2}(u_{1},u_{2})
\notag \\
&+&\left[ \omega _{1}(u)F_{2}(u,u_{1},u_{2},u_{3})+\omega
_{2}(u)G_{2}(u,u_{1},u_{2},u_{3})\right] \mathcal{B}(u)\Psi _{2}(u_{1},u_{3})
\notag \\
&+&\left[ \omega _{1}(u)F_{1}(u,u_{1},u_{2},u_{3})+\omega
_{2}(u)G_{1}(u,u_{1},u_{2},u_{3})\right] \mathcal{B}(u)\Psi
_{2}(u_{2},u_{3}),  \notag \\
&&
\end{eqnarray}%
with 
\begin{equation}
\Lambda _{3}(u,u_{1},u_{2},u_{3})=\omega
_{1}(u)\Delta_1(u)\prod_{j=1}^{3}a_{1}(u,u_{j})+\omega
_{2}(u)\Delta_2(u)\prod_{j=1}^{3}b_{1}(u,u_{j}).  \label{lam3}
\end{equation}

The next step consists in the determination of $t(u)\Phi
_{3}(u_{1},u_{2},u_{3})$. We have a proliferation of cumbersome unwanted
terms in this case and for this reason we omit their expressions. Following
the last subsections, we impose the vanishing of the unwanted terms to fix
relations for the unknowns $g^{(k)}(u_{1},u_{2},u_{3})$ as well as to obtain
the Bethe equations. Three of the unwanted terms in $t(u)\Phi
_{3}(u_{1},u_{2},u_{3})$ coincide with those in (\ref{tPSI3}) and lead to
the Bethe equations, 
\begin{equation}
\frac{\Delta _{1}(u_{k})}{\Delta _{2}(u_{k})}=-\Theta
(u_{k})\prod_{j=1,j\neq k}^{3}{\frac{b_{1}(u_{k},u_{j})}{a_{1}(u_{k},u_{j})}}%
,\quad k=1,2,3  \label{bethe3particle}
\end{equation}%
The $g^{(k)}(u_{1},u_{2},u_{3})$ are obtained from the coefficients of $\Psi
_{2}(u_{i},u_{j})$ $(i<j)$, $\Psi _{1}(u_{i})$ $(i=1,2,3)$ and $\Psi _{0}$.
Once again, taking into account the Bethe equations (\ref{bethe3particle}),
we get 
\begin{eqnarray}
g_{3}^{(2)}(u_{1},u_{2},u_{3}) &=&g(u_{3})p(u_{1},u_{3})p(u_{2},u_{3}) 
\notag \\
g_{2}^{(2)}(u_{1},u_{2},u_{3}) &=&g(u_{2})p(u_{1},u_{2})p(u_{3},u_{2}) 
\notag \\
g_{1}^{(2)}(u_{1},u_{2},u_{3}) &=&g(u_{1})p(u_{2},u_{1})p(u_{3},u_{1})
\label{g3.1}
\end{eqnarray}%
as the factors of the three $\Psi _{2}$, 
\begin{eqnarray}
g_{23}^{(1)}(u_{1},u_{2},u_{3})
&=&g(u_{2})g(u_{3})p(u_{1},u_{2})p(u_{1},u_{3})q(u_{2},u_{3})  \notag \\
g_{13}^{(1)}(u_{1},u_{2},u_{3})
&=&g(u_{1})g(u_{3})p(u_{2},u_{1})p(u_{2},u_{3})q(u_{1},u_{3})  \notag \\
g_{12}^{(1)}(u_{1},u_{2},u_{3})
&=&g(u_{1})g(u_{2})p(u_{3},u_{2})p(u_{3},u_{1})q(u_{1},u_{2})  \label{g3.2}
\end{eqnarray}%
as the factors of the three $\Psi _{1}$ and%
\begin{equation}
g_{123}^{(0)}(u_{1},u_{2},u_{3})=g(u_{1})g(u_{2})g(u_{3})q(u_{1},u_{2})q(u_{1},u_{3})q(u_{2},u_{3})
\label{g3.3}
\end{equation}%
as the factor of $\Psi _{0}$. Here we notice that the expressions for $%
g^{(k)}(u_{1},u_{2},u_{3})$ are factorized in terms of the previously
defined functions $g(u)$, $p(u,v)$ and $q(u,v)$.

Therefore, (\ref{psi3}) is eigenstate of $t(u)$ with eigenvalue (\ref{lam3})
provided the equations (\ref{bethe3particle}) are valid.

\subsection{The general excited state}

The previous results allow us to write the general expressions for the
eigenvalue problem of the transfer matrix (\ref{transfer1}): the nth excited
eigenstate is given by 
\begin{eqnarray}
&&\Phi _{n}(u_{1},\dots ,u_{n})=\Psi_n(u_{1},\dots ,u_{n})  \notag \\
&+& \sum_{k=0}^{n-1} \sum_{\ell _{1}<\dots <\ell_{n-k}=1}^{n} g^{(k)}_{\ell
_{1},\dots ,\ell _{n-k}}(u_{1},\dots ,u_{n}) \Psi _{k}(u_{1},\ldots ,\hat{u}%
_{\ell _{1}},\ldots ,\hat{u}_{\ell _{n-k}},\ldots ,u_{n}),  \notag \\
\end{eqnarray}
where the functions $g^{(k)}_{\ell _{1},\dots ,\ell _{n-k}}(u_{1},\dots
,u_{n})$ have the following expression, 
\begin{equation}
g^{(k)}_{\ell _{1},\dots ,\ell _{n-k}}(u_{1},\dots ,u_{n})= \prod_{m\in\bar{%
\ell}} g(u_{m}) \prod_{m^{\prime }\in\bar{\ell},m^{\prime }<m}
q(u_{m^{\prime }},u_{m}) \prod_{m^{\prime \prime }=1,m^{\prime \prime
}\notin \bar{\ell}}^{n} p(u_{m^{\prime\prime}},u_{m})
\end{equation}
with $\bar{\ell}=\{\ell _{1},\dots ,\ell _{n-k}\}$ and the notation $\hat{u}%
_{j}$ indicates the absence of the rapidity $u_{j}$ in the function. The
corresponding eigenvalue is given by, 
\begin{equation}
\Lambda _{n}(u,u_{1},\ldots ,u_{n})=\omega
_{1}(u)\Delta_1(u)\prod_{j=1}^{n}a_{1}(u,u_{j})+\omega
_{2}(u)\Delta_2(u)\prod_{j=1}^{n}b_{1}(u,u_{j})  \label{lamn}
\end{equation}%
while the Bethe rapidities are constrained by 
\begin{equation}
\frac{\Delta _{1}(u_{k})}{\Delta _{2}(u_{k})}=-\Theta
(u_{k})\prod_{j=1,j\neq k}^{n}{\frac{b_{1}(u_{k},u_{j})}{a_{1}(u_{k},u_{j})}}%
,  \label{bethenparticle}
\end{equation}
where $k=1,\ldots,n$.

\section{Conclusion}

\label{conclu}

We have solved the six vertex model for upper triangular reflection $K$%
-matrices by means of the algebraic Bethe ansatz. The eigenvalues and the
Bethe equations are found to be independent of the upper boundary constants.
However, the Bethe states are essentially different. In fact, the
wavefunctions of the transfer matrix are a superposition of $2^{n}$ Bethe
states of an auxiliary diagonal transfer matrix. This fact may indicate, for
example, the existence of generalized solutions of the
Knizhnik-Zamolodchikov equations, inspired by the semiclassical limit of our
solution \cite{Hikami, ALSUtiel}.

Finally, we remark that our strategy to deal with transfer matrices
possessing annihilation operators in their expression, more than a
particular boundary configuration of the six vertex model, may allow the
management of the generic boundary case, for instance, attempting to extend
the works \cite{cao,Doikou,CRM}. Further directions of investigation include
vertex models based on higher hank algebras, \textit{e.g.}, $15$- or $19$%
-vertex models.

\section{Acknowledgments}

It is ALS's pleasure to thank Professor Roland K\"{o}berle for interesting
discussions. The work of RAP has been supported by S\~{a}o Paulo Research
Foundation (FAPESP), grant \#2012/13126-0. ALS also thanks Brazilian
Research Council (CNPq), grant \#304054/ 2009-7 and FAPESP, grant
\#2011/18729-1 for financial support.

\appendix

\section{Coefficients of the commutation relations}

\label{commutcoef}

The coefficients of the commutation relation (\ref{commutAB}) are given by 
\begin{eqnarray}
a_{1}(u,v) &=&\frac{\sinh (u+v)\sinh (u-v-\eta )}{\sinh (u-v)\sinh (u+v+\eta
)},~~a_{2}(u,v)=\frac{\sinh (2v)\sinh (\eta )}{\sinh (u-v)\sinh (2v+\eta )},
\notag  \label{commutABcoef} \\
a_{3}(u,v) &=&-\frac{\sinh (\eta )}{\sinh (u+v+\eta )},
\end{eqnarray}%
and the coefficients of (\ref{commutDB}) are%
\begin{eqnarray}
b_{1}(u,v) &=&\frac{\sinh (u-v+\eta )\sinh (u+v+2\eta )}{\sinh (u-v)\sinh
(u+v+\eta )},~~b_{2}(u,v)=\frac{\sinh (\eta )\sinh [2(u+\eta )]}{\sinh
(v-u)\sinh (2u+\eta )},  \notag  \label{commutDBcoef} \\
b_{3}(u,v) &=&\frac{\sinh (2v)\sinh (\eta )\sinh [2(u+\eta )]}{\sinh
(2u+\eta )\sinh (2v+\eta )\sinh (u+v+\eta )},
\end{eqnarray}%
while the coefficients of (\ref{commutCB}) are given by 
\begin{eqnarray}
c_{1}(u,v) &=&1,~~c_{2}(u,v)=\frac{\sinh (2u)\sinh (\eta )\sinh (u-v+\eta )}{%
\sinh (u-v)\sinh (2u+\eta )\sinh (u+v+\eta )},  \notag  \label{commutCBcoef}
\\
c_{3}(u,v) &=&\frac{\sinh (2u)\sinh ^{2}(\eta )}{\sinh (v-u)\sinh (2u+\eta
)\sinh (2v+\eta )},  \notag \\
c_{4}(u,v) &=&\frac{\sinh (u+v)\sinh (\eta )}{\sinh (u-v)\sinh (u+v+\eta )}%
,~~c_{5}(u,v)=\frac{\sinh (2u)\sinh (\eta )}{\sinh (v-u)\sinh (2u+\eta )}, 
\notag \\
c_{6}(u,v) &=&-\frac{\sinh ^{2}(\eta )}{\sinh (u+v+\eta )\sinh (2v+\eta )}%
,~~c_{7}(u,v)=-\frac{\sinh (\eta )}{\sinh (u+v+\eta )}.
\end{eqnarray}

The respective rational coefficients are obtained by the substitution $\sinh
(x)\rightarrow x$ in the above expressions.

\section{Reordered operators}

\label{reorder}

An important point in the {\small ABA} analysis is to move the operators $%
\mathcal{A}(u)$, $\tilde{\mathcal{D}}(u)$ and $\mathcal{C}(u)$ over the
product $\prod_{j=1}^{n}\mathcal{B}(u_{j})\Psi _{0}$ and then use (\ref%
{actionrefs}). The repeated use of the commutation relations (\ref{commutBB}%
) to (\ref{commutCB}) allow us to write, 
\begin{eqnarray}
\mathcal{A}(u)\prod_{j=1}^{n}\mathcal{B}(u_{j})\Psi _{0} &=&\left[ \Delta
_{1}(u)\prod_{j=1}^{n}a_{1}(u,u_{j})\right] \prod_{j=1}^{n}\mathcal{B}%
(u_{j})\Psi _{0}  \notag  \label{ABn} \\
&+&\sum_{k=1}^{n}F_{k}(u,u_{1},\ldots ,u_{n})\mathcal{B}(u)\prod_{j=1,j\neq
k}^{n}\mathcal{B}(u_{j})\Psi _{0},
\end{eqnarray}%
\begin{eqnarray}
\tilde{\mathcal{D}}(u)\prod_{j=1}^{n}\mathcal{B}(u_{j})\Psi _{0} &=&\left[
\Delta _{2}(u)\prod_{j=1}^{n}b_{1}(u,u_{j})\right] \prod_{j=1}^{n}\mathcal{B}%
(u_{j})\Psi _{0}  \notag  \label{DBn} \\
&+&\sum_{k=1}^{n}G_{k}(u,u_{1},\ldots ,u_{n})\mathcal{B}(u)\prod_{j=1,j\neq
k}^{n}\mathcal{B}(u_{j})\Psi _{0},
\end{eqnarray}%
\begin{eqnarray}
\mathcal{C}(u)\prod_{j=1}^{n}\mathcal{B}(u_{j})\Psi _{0}
&=&\sum_{k=1}^{n}H_{k}(u,u_{1},\ldots ,u_{n})\prod_{j=1,j\neq k}^{n}\mathcal{%
B}(u_{j})\Psi _{0}  \notag  \label{CBn} \\
&+&\sum_{\ell >k}^{n}H_{\ell k}(u,u_{1},\ldots ,u_{n})\mathcal{B}%
(u)\prod_{j=1,j\neq \ell ,k}^{n}\mathcal{B}(u_{j})\Psi _{0},
\end{eqnarray}%
where 
\begin{equation}
F_{k}(u,u_{1},\ldots ,u_{n})=\Delta _{1}(u_{k})a_{2}(u,u_{k})\prod_{\ell
=1,\ell \neq k}^{n}a_{1}(u_{k},u_{\ell })+\Delta
_{2}(u_{k})a_{3}(u,u_{k})\prod_{\ell =1,\ell \neq k}^{n}b_{1}(u_{k},u_{\ell
}),  \label{defFk}
\end{equation}%
\begin{equation}
G_{k}(u,u_{1},\ldots ,u_{n})=\Delta _{1}(u_{k})b_{3}(u,u_{k})\prod_{\ell
=1,\ell \neq k}^{n}a_{1}(u_{k},u_{\ell })+\Delta
_{2}(u_{k})b_{2}(u,u_{k})\prod_{\ell =1,\ell \neq k}^{n}b_{1}(u_{k},u_{\ell
}),  \label{defGk}
\end{equation}%
\begin{eqnarray}
H_{k}(u,u_{1},\ldots ,u_{n}) &=&\Delta _{1}(u)\Delta _{1}(u_{k})\left[
c_{2}(u,u_{k})+c_{3}(u,u_{k})\right] \prod_{\ell=1,\ell\neq
k}^{n}a_{1}(u,u_{\ell})a_{1}(u_{k},u_{\ell})  \notag  \label{defHk} \\
&+&\Delta _{2}(u)\Delta _{1}(u_{k})\left[ c_{4}(u,u_{k})+c_{6}(u,u_{k})%
\right] \prod_{\ell=1,\ell\neq k}^{n}b_{1}(u,u_{\ell})a_{1}(u_{k},u_{\ell}) 
\notag \\
&+&\Delta _{1}(u)\Delta _{2}(u_{k})c_{5}(u,u_{k})\prod_{\ell=1,\ell\neq
k}^{n}a_{1}(u,u_{\ell})b_{1}(u_{k},u_{\ell})  \notag \\
&+&\Delta _{2}(u)\Delta _{2}(u_{k})c_{7}(u,u_{k})\prod_{\ell=1,\ell\neq
k}^{n}b_{1}(u,u_{\ell})b_{1}(u_{k},u_{\ell}),
\end{eqnarray}%
\begin{eqnarray}
H_{\ell k}(u,u_{1},\ldots ,u_{n}) &=& \Delta _{1}(u_{k})\Delta
_{1}(u_{\ell})\alpha _{11}(u,u_{k},u_{\ell }) \prod_{m=1,m\neq
\ell,k}^{n}a_{1}(u_{k},u_{m})a_{1}(u_{\ell },u_{m})  \notag  \label{defHlk}
\\
&+& \Delta _{1}(u_{k})\Delta _{2}(u_{\ell })\alpha _{12}(u,u_{k},u_{\ell})
\prod_{m=1,m\neq \ell ,k}^{n}a_{1}(u_{k},u_{m})b_{1}(u_{\ell },u_{m})  \notag
\\
&+& \Delta _{1}(u_{\ell })\Delta _{2}(u_{k})\alpha _{21}(u,u_{k},u_{\ell})
\prod_{m=1,m\neq \ell ,k}^{n}a_{1}(u_{\ell },u_{m})b_{1}(u_{k},u_{m})  \notag
\\
&+& \Delta _{2}(u_{\ell })\Delta _{2}(u_{k})\alpha _{22}(u,u_{k},u_{\ell})
\prod_{m=1,m\neq \ell ,k}^{n}b_{1}(u_{\ell },u_{m})b_{1}(u_{k},u_{m}),
\end{eqnarray}%
with 
\begin{eqnarray}
\alpha _{11}(u,u_{k},u_{\ell}) &=&a_{2}(u,u_{\ell})\left[
a_{1}(u_{k},u)c_{2}(u,u_{k})+c_{3}(u,u_{k})a_{1}(u_{k},u_{\ell})\right] 
\notag \\
&+&b_{3}(u,u_{\ell})\left[
a_{1}(u_{k},u)c_{4}(u,u_{k})+c_{6}(u,u_{k})a_{1}(u_{k},u_{\ell})\right] 
\notag \\
&+&a_{2}(u,u_{k})\left[ c_{3}(u,u_{k})a_{2}(u_{k},u_{\ell})
+c_{5}(u,u_{k})b_{3}(u_{k},u_{\ell})\right]  \notag \\
&+&b_{3}(u,u_{k})\left[ c_{6}(u,u_{k})a_{2}(u_{k},u_{\ell})+
c_{7}(u,u_{k})b_{3}(u_{k},u_{\ell})\right] ,
\end{eqnarray}%
\begin{eqnarray}
\alpha _{12}(u,u_{k},u_{\ell}) &=&a_{3}(u,u_{\ell})\left[
a_{1}(u_{k},u)c_{2}(u,u_{k})+c_{3}(u,u_{k})a_{1}(u_{k},u_{\ell })\right] 
\notag \\
&+&b_{2}(u,u_{\ell })\left[
a_{1}(u_{k},u)c_{4}(u,u_{k})+c_{6}(u,u_{k})a_{1}(u_{k},u_{\ell })\right] 
\notag \\
&+&a_{2}(u,u_{k})\left[ c_{3}(u,u_{k})a_{3}(u_{k},u_{\ell
})+c_{5}(u,u_{k})b_{2}(u_{k},u_{\ell })\right]  \notag \\
&+&b_{3}(u,u_{k})\left[ c_{6}(u,u_{k})a_{3}(u_{k},u_{\ell
})+c_{7}(u,u_{k})b_{2}(u_{k},u_{\ell })\right] ,
\end{eqnarray}%
\begin{eqnarray}
\alpha _{21}(u,u_{k},u_{\ell}) &=&c_{5}(u,u_{k})\left[ a_{2}(u,u_{\ell
})b_{1}(u_{k},u_{\ell })+a_{3}(u,u_{k})b_{3}(u_{k},u_{\ell })\right]  \notag
\\
&+&c_{7}(u,u_{k})\left[ b_{3}(u,u_{\ell })b_{1}(u_{k},u_{\ell
})+b_{2}(u,u_{k})b_{3}(u_{k},u_{\ell })\right]  \notag \\
&+&a_{2}(u_{k},u_{\ell })\left[
a_{3}(u,u_{k})c_{3}(u,u_{k})+b_{2}(u,u_{k})c_{6}(u,u_{k})\right] ,  \notag \\
&&
\end{eqnarray}%
\begin{eqnarray}
\alpha _{22}(u,u_{k},u_{\ell}) &=&c_{5}(u,u_{k})\left[ a_{3}(u,u_{\ell
})b_{1}(u_{k},u_{\ell })+a_{3}(u,u_{k})b_{2}(u_{k},u_{\ell })\right]  \notag
\\
&+&c_{7}(u,u_{k})\left[ b_{2}(u,u_{\ell })b_{1}(u_{k},u_{\ell
})+b_{2}(u,u_{k})b_{2}(u_{k},u_{\ell })\right]  \notag \\
&+&a_{3}(u_{k},u_{\ell })\left[
a_{3}(u,u_{k})c_{3}(u,u_{k})+b_{2}(u,u_{k})c_{6}(u,u_{k})\right] .  \notag \\
&&
\end{eqnarray}

We note that for diagonal boundaries only the expressions (\ref{ABn}) and (%
\ref{DBn}) are necessary \cite{Skly} while for the upper triangular $K-$%
matrices case we also need the more involved relation (\ref{CBn}).


\begin{thebibliography}{99}
\bibitem{Skly} E.K. Sklyanin, ``Boundary conditions for integrable quantum
systems,'' \textit{J. Phys. A: Math. Gen.} \textbf{21} (1988) 2375

\bibitem{Che} I.V. Cherednik, ``Factorizing particles on a half-line and
root systems,'' \textit{Theo. Math. Phys.} \textbf{61} (1984) 977

\bibitem{Fan} H. Fan, ``Bethe ansatz for the Izergin-Korepin model,'' 
\textit{Nucl. Phys. B} \textbf{488} (1997) 409

\bibitem{Guan} X.-W. Guan, ``Algebraic Bethe ansatz for the one-dimensional
Hubbard model with open boundaries'', \textit{J. Phys. A: Math. Gen.} 
\textbf{33} (2000) 5391

\bibitem{LiYueHou} G.-L. Li, R.-H. Yue and B.-Y. Hou, ``Nested Bethe ansatz
for Perk-Schultz model with open boundary conditions,'' \textit{Nucl. Phys. B%
} \textbf{23} (2000) 711

\bibitem{KurakLima} V. Kurak and A. Lima-Santos, ``Algebraic Bethe ansatz
for the Zamolodchikov-Fateev and Izergin-Korepin models with open boundary
conditions,'' \textit{Nucl. Phys. B} \textbf{699} (2004) 595

\bibitem{LiShi} G.-L. Li and K.-J. Shi, ``The algebraic Bethe ansatz for
open vertex models,'' \textit{J. Stat. Mech.} (2007) P01018

\bibitem{Nepo1} R.I. Nepomechie, ``Solving the open XXZ spin chain with
nondiagonal boundary terms at roots of unity,'' \textit{Nucl. Phys. B} 
\textbf{622} (2002) 615

\bibitem{Nepo2} R.I. Nepomechie, ``Functional relations and Bethe Ansatz for
the XXZ chain,'' \textit{J. Stat. Phys.} \textbf{111} (2003) 1363;

R.I. Nepomechie, ``Bethe ansatz solution of the open XXZ chain with
nondiagonal boundary terms,'' \textit{J. Phys. A: Math. Gen.} \textbf{37}
(2004) 433

\bibitem{MurNepo} R. Murgan and R.I. Nepomechie, ``Bethe Ansatz derived from
the functional relations of the open XXZ chain for new special cases'' 
\textit{J. Stat. Mech.} (2005) P05007;

R. Murgan and R.I. Nepomechie, ``Generalized T-Q relations and the open XXZ
chain,'' \textit{J. Stat. Mech.} (2005) P08002

\bibitem{Nepoetal} W.-L. Yang, R.I. Nepomechie and Y.-Z. Zhang, ``Q-operator
and T-Q relation from the fusion hierarchy,'' \textit{Phys. Lett. B} \textbf{%
633} (2006) 664;

R. Murgan, R.I. Nepomechie and C. Shi, ``Exact solution of the open XXZ
chain with general integrable boundary terms at roots of unity,'' \textit{J.
Stat. Mech.} (2006) P08006;

L. Frappat, R.I. Nepomechie and E. Ragoucy, ``A complete Bethe ansatz
solution for the open spin-s XXZ chain with general integrable boundary
terms,'' \textit{J. Stat. Mech.} (2007) P09009

\bibitem{CYSW} J. Cao, W.-L. Yang, K. Shi, and Y. Wang, ``Off-diagonal Bethe
ansatz solution of the XXX spin-chain with arbitrary boundary conditions,'' 
\textit{Nucl. Phys. B} \textbf{875} (2013) 152;

J. Cao, W. Yang, K. Shi, and Y. Wang, ``Off-diagonal Bethe ansatz and exact
solution a topological spin ring,'' \texttt{arXiv:1305.7328};

J. Cao, W.-L. Yang, K. Shi, and Y. Wang, ``Off-diagonal Bethe ansatz
solutions of the anisotropic spin-1/2 chains with arbitrary boundary
fields,'' \texttt{arXiv:1307.2023}

\bibitem{Nepo2013} R.I. Nepomechie, ``Inhomogeneous T-Q equation for the
open XXX chain with general boundary terms: completeness and arbitrary
spin,'' \texttt{arXiv:1307.5049}

\bibitem{FGSW} H. Frahm, A. Seel and T. Wirth, ``Separation of variables in
the open XXX chain,'' \textit{Nucl. Phys. B} \textbf{802} (2008) 351;

H. Frahm, J.H. Grelik, A. Seel and T. Wirth, ``Functional Bethe ansatz
methods for the open XXX chain,'' \textit{J. Phys. A: Math. Theor.} \textbf{%
44} (2011) 015001

\bibitem{NFK} G. Niccoli, ``Non-diagonal open spin-1/2 XXZ quantum chains by
separation of variables: Complete spectrum and matrix elements of some
quasi-local operators,'' \textit{J. Stat. Mech.} (2012) P10025;

S. Faldella, N. Kitanine and G. Niccoli, ``Complete spectrum and scalar
products for the open spin-1/2 XXZ quantum chains with non-diagonal boundary
terms,'' \texttt{arXiv:1307.3960}

\bibitem{Gal} W. Galleas, ``Functional relations from the Yang-Baxter
algebra: Eigenvalues of the XXZ model with non-diagonal twisted and open
boundary conditions,'' \textit{Nucl. Phys. B} \textbf{790} (2008) 524

\bibitem{BK} P. Baseilhac and K. Koizumi, ``Exact spectrum of the XXZ open
spin chain from the q-Onsager algebra representation theory,'' \textit{J.
Stat. Mech.} (2007) P09006;

\bibitem{cao} J. Cao, H.-Q. Lin, K.-J. Shi and Y. Wang, ``Exact solutions
and elementary excitations in the XXZ spin chain with unparallel boundary
fields,'' \texttt{arXiv:0212163};

J. Cao, H.-Q. Lin, K.-J. Shi and Y. Wang, ``Exact solution of XXZ spin chain
with unparallel boundary fields,'' \textit{Nucl. Phys. B} \textbf{663}
(2003) 487

\bibitem{Doikou} A. Doikou, ``A note on the boundary spin s XXZ chain,'' 
\textit{Phys. Lett. A} \textbf{366} (2007) 556

\bibitem{CRM} C.S. Melo, G.A.P. Ribeiro and M.J. Martins, ``Bethe ansatz for
the XXX-S chain with non-diagonal open boundaries,'' \textit{Nucl. Phys. B} 
\textbf{711} (2005) 565

\bibitem{BCR} S. Belliard, N. Cramp\'e and E. Ragoucy, ``Algebraic Bethe
Ansatz for Open XXX Model with Triangular Boundary Matrices,'' \textit{Lett.
Math. Phys.} \textbf{103} (2013) 493

\bibitem{CR} N. Cramp\'e and E. Ragoucy, ``Generalized coordinate Bethe
ansatz for non diagonal boundaries,'' \textit{Nucl. Phys. B} \textbf{858}
(2012) 502

\bibitem{BB} P. Baseilhac and S. Belliard, ``The half-infinite XXZ chain in
Onsager's approach,'' \textit{Nucl. Phys. B} \textbf{873}
(2013) 550

\bibitem{VegaRuiz} H.J. de Vega and A.G. Ruiz, ``Boundary K-matrices for the
six vertex and the $n(2n-1)A_{n-1}$ vertex models,'' \textit{J. Phys. A:
Math. Gen.} \textbf{26} (1993) L519

\bibitem{GhoZamo} S. Ghoshal and A. Zamolodchikov, ``Boundary S-Matrix and
Boundary State in Two-Dimensional Integrable Quantum Field Theory,'' \textit{%
Int. J. Mod. Phys. A} \textbf{9} (1994) 3841

\bibitem{Hikami} K. Hikami, ``Gaudin magnet with boundary and generalized
Knizhnik-Zamolodchikov equation ,'' \textit{J. Phys. A: Math. Gen.} \textbf{%
28} (1995) 4997

\bibitem{ALSUtiel} A. Lima-Santos and W. Utiel, ``Gaudin magnet with
impurity and its generalized Knizhnik-Zamolodchikov Equation," \textit{Int.
J. Mod. Phys. B} \textbf{20} (2006) 2175
\end{thebibliography}
\end{document}